\documentclass[12pt,eadjoint tfnpsf]{article}

\catcode`\@=11
\@addtoreset{equation}{section}

\global\arraycolsep=1pt
\usepackage{amsbsy,amssymb,latexsym,amsfonts}
\usepackage{mathrsfs}
\usepackage{graphicx}

\RequirePackage[dvips,usenames]{color}
\definecolor{fireblick}{rgb}{0.698039,0.133333,0.133333}

\newcommand{\beq}{\begin{equation}}
\newcommand{\eeq}{\end{equation}}
\newcommand{\bea}{\begin{eqnarray}}
\newcommand{\eea}{\end{eqnarray}}

\newcommand{\CD}{{\mathcal D}}
\newcommand{\CF}{{\mathcal F}}
\newcommand{\CL}{{\mathcal L}}
\newcommand{\CN}{{\mathcal N}}

\newcommand{\CW}{{\mathcal W}}

\renewcommand\Im{{\mathrm{Im}}}

\newcommand\ui{{\underline{i}}}

\def\Tr{\mathop{\rm Tr}}

\begin{document}
%
%
\begin{titlepage}

\begin{flushright}
\normalsize
~~~~
August, 2008 \\
OCU-PHYS 303 \\
\end{flushright}

\vspace{36pt}

\begin{center}
{\LARGE SUSY/Non-SUSY Duality in $U(N)$ Gauge Model}\\
\vspace{11pt}
{\LARGE with Partially Broken $\CN=2$ Supersymmetry}
\end{center}

\vspace{36pt}

\begin{center}
{%
Kazunobu Maruyoshi\footnote{e-mail: maruchan@sci.osaka-cu.ac.jp}
}\\
%
\vspace{12pt}
%
\it Department of Mathematics and Physics, Graduate School of Science\\
Osaka City University\\
\vspace{12pt}
3-3-138, Sugimoto, Sumiyoshi-ku, Osaka, 558-8585, Japan \\
\end{center}
%
\vspace{36pt}
%
  We study the vacuum structure of the $U(N)$ gauge model with partially broken $\CN=2$ supersymmetry.
  From the analysis of the classical vacua of this model, 
  we point out that in addition to the ordinary $\CN=1$ supersymmetric vacua, 
  there are vacua with negative gauge coupling constants, which preserve another $\CN=1$ supersymmetry.
  These latter vacua can be analyzed by using SUSY/non-SUSY duality 
  which is recently proposed by Aganagic, Beem, Seo and Vafa.
  A dual description of these in UV is $U(N)$ gauge theory where the supersymmetry is broken by spurion superfields.
  Following them, we see that there are supersymmetry preserving vacua as well as supersymmetry breaking vacua 
  of low energy effective theory.

\vfill

\setcounter{footnote}{0}
\renewcommand{\thefootnote}{\arabic{footnote}}

\end{titlepage}

\section{Introduction}
\label{sec:intro}
  The low energy behavior of $\CN=1$ supersymmetric gauge theory has been studied in various contexts 
  and a lot of exciting results have been found.
  In the last decade, various investigations have been made on
  the effective superpotential of $U(N)$ super Yang-Mills theory with an adjoint chiral superfield 
  and a tree level superpotential, which is the single trace function of the adjoint chiral superfield. 
  It has been known that this can be computed 
  by using gauge/gravity correspondence in superstring theory \cite{Vafa, CIV, CV}, 
  by the bosonic one matrix model \cite{DV},
  more directly by its Riemann surface data \cite{IMorozov}
  and also by purely field theoretical methods \cite{DGLVZ, CDSW, CSW}.
  
  Recently, the study of the effective superpotential has entered a new phase.
  In \cite{IM, Ferrari, IM2}, it has been shown that,
  in the case where a gauge kinetic term depends on the adjoint chiral superfield, 
  the effective superpotential contains a term which deforms the form of the one obtained by \cite{Vafa, CIV, DV}.
  More precisely, the model analyzed in \cite{Ferrari} has the following holomorphic terms in classical Lagrangian
    \bea
    \int d^2 \theta \Tr \left[ \alpha(\Phi) \CW^\alpha \CW_\alpha + W(\Phi) \right],
    ~~
    \alpha (\Phi) 
     =     \sum_{k=0}^{n-1} t_k \Phi^k,
           ~~
    W (\Phi) 
     =     \sum_{k=1}^{n+1} a_k \Phi^k,
    \label{genericholo}
    \eea
  where $\alpha(\Phi)$ and a tree level superpotential
  are the quite generic polynomials of the chiral superfield and are not related each other.
  We refer to this model as generic model.
  In contrast, the $U(N)$ gauge model \cite{FIS, FIS2} (FIS)
  whose gauge kinetic term and superpotential are expressed by a bare prepotential has been analyzed in \cite{IM, IM2}
  and the deformed effective superpotential has been derived.
  While it may seem that this latter model is merely one case of the former generic model, 
  it is a special case and has higher symmetry than that of the generic model: 
  the model (FIS) has $\CN=2$ supersymmetry which is spontaneously broken to $\CN=1$.
  This partial breaking model is the non-abelian generalization of the models considered in \cite{APT, FGP, IZ, TV}.
  (See also \cite{FIS3,IMS} for the case with hypermultiplets, 
  \cite{sugra} for $\CN=2$ supergravity and \cite{KMG} for related discussions.)
  
  One consequence which can be argued from the deformation of the effective superpotential is 
  the existence of supersymmetry breaking vacua in such theories.
  The reason is as follows: in the case without deformation, 
  the effective superpotential takes the same form as in \cite{TV, Vafa, ABSV0}.
  This effective theory has two types of vacua: 
  the one preserves an original $\CN=1$ supersymmetry 
  and the other preserves another $\CN=1$ supersymmetry which is a part of $\CN=2$ supersymmetry 
  in the low energy effective theory.
  If there is a deformation, the situation becomes quite different.
  The effective superpotential is no longer the form of \cite{TV, Vafa, ABSV0}
  and the latter vacua become supersymmetry breaking vacua, like in \cite{OOP, HMPS}.
  
  These supersymmetric vacua and supersymmetry breaking vacua do not exist at the same time, however.
  To be accurate, for the choices of the parameters where the classical theory has positive kinetic energy: 
  $(1/g^2)_i \sim \langle \Im \alpha \rangle_i \gg 0$ for all $i$ 
  (indices $i$ label $U(N_i)$ gauge groups which are unbroken in the classical vacua), 
  only the supersymmetric vacua are in the region where the effective theory is valid 
  and the supersymmetry breaking vacua are out of this region.
  On the other hand, for the parameters where the squared gauge coupling constants are negative, 
  the supersymmetry breaking vacua are valid 
  and the supersymmetric vacua are out of the region of validity of the effective theory.
  In this case, the field theory description in UV is bad.
  
  Based on these observations, an interesting duality has proposed \cite{ABSV} in the generic model:
  in the case when all the squared gauge coupling constants are negative, 
  there exists a good field theory description in UV,
  which is $U(N)$ gauge model where the squared gauge coupling constants are positive 
  and has the following superpotential 
    \bea
    \widetilde{W}(\widetilde{\Phi})
     =     \sum_{k=1}^{n+1} (a_k + 2 i t_k \tilde{\theta}^2) \Phi^k.
    \eea
  This is the model with spurion fields 
  which has nonzero vevs in auxiliary fields and they break the supersymmetry explicitly \cite{LM, GMT}.
  Also, this duality has been analyzed from type IIA superstring and M-theory perspective in \cite{OT}.
  
  The model we study in this paper is the $U(N)$ gauge model (FIS) 
  where classically $\CN=2$ supersymmetry is broken to $\CN=1$.
  This is the specific case of the generic model.
  Therefore, we have ordinary $\CN=1$ supersymmetric vacua and, 
  following the above discussion of SUSY/non-SUSY duality, 
  there exists the supersymmetry breaking vacua in the case 
  where classically the squared gauge coupling constants are negative.
  
  We notice that there are other supersymmetric vacua in the classical theory, 
  which preserve different $\CN=1$ supersymmetry from the one preserved in the above vacua.
  This is due to the $\CN=2$ structure of this model and what we focus on in this paper.
  We can see that 
  in the case where the squared gauge coupling constants are positive in the ordinary $\CN=1$ supersymmetric vacua,
  they are negative in the second $\CN=1$ supersymmetric vacua, and vice versa.
  We then analyze the second $\CN=1$ supersymmetric vacua where the squared gauge coupling constants are negative 
  by applying the above duality
  and see that these lead to the supersymmetry breaking vacua at the low energy.
  Therefore, we will see the existence of 
  the supersymmetry breaking vacua as well as the supersymmetric vacua in this model.
  From the above analysis, one may observe the similarity between $\CN=2$ structure 
  of the classical theory and that of the effective theory.
  However, we see that these are not exactly same, by considering the limit 
  where the model reduces to $\CN=1$, $U(N)$ super Yang-Mills theory with the superpotential.
  
  The organization of this paper is as follows. 
  We review SUSY/non-SUSY duality in the above generic model in section \ref{sec:duality}.
  After that, we concentrate on the $U(N)$ gauge model with partially broken $\CN=2$ supersymmetry
  in the subsequent sections.
  In section \ref{sec:model}, we extend the analysis of the classical vacua and its classification of \cite{FIS2}
  in order to find out the consequence from $\CN=2$ supersymmetry in the original Lagrangian.
  We also see the unbroken supersymmetry in these vacua.
  In section \ref{sec:non-susy}, the effective superpotential of this model will be considered 
  and we will see that both the supersymmetric vacua and the supersymmetry breaking vacua
  can be analyzed by using SUSY/non-SUSY duality.
  Finally, we conclude in section \ref{sec:summary}.

\section{SUSY/Non-SUSY Duality}
\label{sec:duality}
  In this section, we briefly see the duality which was proposed in \cite{ABSV} in the generic model (\ref{genericholo}).
  First of all, let us consider the region of the parameters $a_k$ and $t_k$, 
  such that all the square inverses of the gauge coupling constants are positive:
    \bea
    \left( \frac{4 \pi}{g^2} \right)_i 
     =     \Im \alpha (\langle \phi^i \rangle) 
     \gg     0,
           \label{>}
    \eea
  where $\langle \phi^i \rangle$ are the diagonal components of the vev of the scalar field 
  and are derived from F-term condition.
  According to the vev, the gauge symmetry is broken to $\prod_{i=1}^n U(N_i)$.
  Each value of (\ref{>}) denotes the gauge coupling constant of each gauge group $U(N_i)$.
  
  In this region, the effective superpotential can be evaluated by purely field theoretical method \cite{Ferrari}
  and also by using the gauge/gravity correspondence and analyzing dual gravity theory with flux \cite{ABSV}.
  The answer is
    \bea
    \int d^2 \theta W_{{\rm eff}}(S_i),
    ~~~
    W_{{\rm eff}}
     =     N_i \frac{\partial F_{{\rm free}}}{\partial S_i}
         + \sum_i \alpha(\langle \phi^i \rangle) S_i
         + \sum_{k>0} t_k \frac{\partial F_{{\rm free}}}{\partial a_k},
           \label{weffgeneric}
    \eea
  where $F_{{\rm free}}$ is the free energy of the bosonic one matrix model 
  whose action is the same form as the tree level superpotential of the gauge theory.
  The first and second terms are usual form which has derived in \cite{Vafa, CIV, CV, DV}:
  the first one is due to the tree level superpotential 
  and the second one comes from $\sum_i \alpha(\langle \phi^i \rangle) \CW^{i \alpha} \CW^i_\alpha$ terms 
  in the classical Lagrangian.
  In addition, there is a deformation term which is the last one in (\ref{weffgeneric}).
  The existence of this is the crucial point of this type of model 
  where a gauge kinetic term depends on the adjoint chiral superfield.
  
  The supersymmetric condition is derived from $\partial_{S_i} W_{{\rm eff}} = 0$.
  From this, we can determine the vevs of the condensate fields.
  But, these vevs should be in the range where the effective theory is valid, which is 
  $\langle S_i \rangle \ll |m \Lambda^2_0|$
    \footnote{We have assumed that the masses of the adjoint chiral superfields of $U(N_i)$ gauge groups
              are the same order and write them as $m$}.
  In the case of (\ref{>}), we can verify that the vevs of the condensate fields are in the above range \cite{ABSV}.
  Furthermore, as noticed in the introduction, we can also see the existence of non-supersymmetric vacua 
  by considering the extremal of the scalar potential.
  However, these non-supersymmetric vacua are not in the range of validity of the effective theory 
  in these parameters region where (\ref{>}) is satisfied.
  Therefore, the meaningful vacua are only the supersymmetric vacua.

  Then, we try to tune the parameters, in the effective superpotential, to the region 
  where all the gauge couplings in the classical theory are negative:
    \bea
    \left( \frac{4 \pi}{g^2} \right)_i 
     =     \Im \alpha (e_i) 
     <     0.
    \eea
  In this choice of the parameters, the supersymmetry breaking vacua are in the region 
  where the effective theory is valid.
  The proposal in \cite{ABSV} is that the effective superpotential in this new choice of the parameters 
  should be considered as that of a better description in UV, 
  rather than the generic model with negative kinetic energy classically.
  This better high energy description is $U(N)$ super Yang-Mills theory 
  with the following superpotential 
    \bea
    \int d^2 \tilde{\theta} \widetilde{W}(\widetilde{\Phi}),
    ~~~
    \widetilde{W}(\widetilde{\Phi})
     =     \sum_{k=1}^{n+1} (a_k + 2 i t_k \tilde{\theta}^2) \Phi^k,
    \eea
  where $\tilde{\theta}$ is a different superspace from the original superspace $\theta$ 
  in the Lagrangian (\ref{genericholo}).
  This is the model with spurion fields 
  which has nonzero vevs in auxiliary fields and they break the supersymmetry explicitly \cite{LM, GMT}.
  
\section{$U(N)$ Gauge Model with Partially Broken $\CN=2$ Supersymmetry}
\label{sec:model}
  From now on, we will consider the specific case of the above generic model, 
  which is $U(N)$ gauge model with partially broken $\CN=2$ supersymmetry.
  If we appropriately choose $t_k$ and $a_k$ in the generic model, we can obtain the following Lagrangian:
    \bea
    \CL
     =     \Im 
           \left[ 
           \int d^4 \theta 
           {\rm Tr} 
           \bar{\Phi} e^{ad V} 
           \frac{\partial \CF(\Phi)}{\partial \Phi}
         + \int d^2 \theta
           \frac{1}{2} 
           \frac{\partial^2 \CF(\Phi)}{\partial \Phi^a \partial \Phi^b}
           \CW^{\alpha a} \CW^b_{\alpha}
           \right]
         + \int d^2 \theta \Tr W(\Phi)
         + h.c.,      
           \label{lagrangian}
    \eea
  and
    \bea
    W(\Phi)
     =     2 e \Phi
         + m \frac{\partial \CF(\Phi)}{\partial \Phi},
           ~~~
    \CF(\Phi)
     =     \sum_{k=1}^{n+1} \frac{g_k}{(k+1)!} \Tr \Phi^{k+1}.
           \label{prepot}
    \eea
  where $m$ is a real parameter and $e$ and $g_k$ ($k=1,\ldots,n+1$) are complex parameters.
  In terms of $U(N)$ generators $t_a$, $a= 0, \ldots, N^2 -1$ ($a = 0$ refers to the overall $U(1)$ generator), 
  the superfield $\Psi = \{ V, \Phi \}$ is $\Psi = \Psi^a t_a$.
  (We normalize the generators as $\Tr (t_a t_b) = \delta_{ab}/2$.)
  As explained in \cite{FIS, FIS2}, this Lagrangian has $\CN=2$ supersymmetry.
  In fact, when $e$ and $m$ are equal to zero, 
  the above Lagrangian is the usual form of $\CN=2$ supersymmetric Lagrangian in $\CN=1$ superspace formalism.
  Non-zero $e$ and $m$ terms, respectively, correspond to the electric and magnetic Fayet-Iliopoulos (FI) terms 
  from the perspective of $\CN=2$ superspace formalism.
  In general, the electric and magnetic FI parameters,
  which are the coefficients of electric and magnetic FI terms, are vectors in $SU(2)_R$.
  We can fix these parameters by using $SU(2)_R$ rotation.
  The parameters $e$ and $m$ in (\ref{lagrangian}) are non-zero components of these vectors after fixing \cite{FIS3,IMS}.
  
  Note that this Lagrangian is slightly different from that in \cite{FIS, FIS2, IM} 
  where there exists a Fayet-Iliopoulos D-term and the parameter $e$ in the superpotential is real.
  Instead, the FI D-term does not exist and $e$ is a complex parameter here. 
  However, from the $\CN=2$ point of view, 
  this is the only difference of the way of fixing the $\CN=2$ electric Fayet-Iliopoulos parameter.
  This does not change the properties of the model as we will see below.
  
  In this section, we will extend the analysis \cite{FIS2} of the classical vacua of this model. 
  Then, we will see the unbroken supersymmetry in these vacua.
  Finally, we explain the limit where the model reduces to $\CN=1$, $U(N)$ super Yang-Mills theory 
  with the superpotential considered by \cite{Vafa, CIV, CV, DV}.

\subsection{Classical vacua of the model}
\label{subsec:vacua}
  While this action is shown to be invariant under the $\CN=2$ supersymmetry
  transformations, the vacuum breaks half of the $\CN=2$ supersymmetry \cite{FIS, FIS2}.
  That is, the vacua of this model preserve only $\CN=1$ supersymmetry.
  This is the property of this model.
  First of all, let us consider the vacua and classify them.
  
  We can easily see supersymmetric vacua from the F-term condition $\partial_a W (\Phi) = 0$.
  Of course, these correspond to the $\CN=1$ supersymmetric vacua which discussed in the generic model.
  There are other vacua which preserve different $\CN=1$ supersymmetry in this model.
  In order to see this, we will analyze the scalar potential:
    \bea
    V
     =     g^{ab}
           \left(
           \frac{1}{8} \CD_a \CD_b 
         + \partial_a  W (\Phi)
           \overline{
           \partial_{b} W (\Phi)}
           \right), 
           \label{potential}
    \eea
  where $\CD_a = - i g_{ab} f^b_{cd} \bar{\phi}^c \phi^d$.
  We are interested in the vacua where $\langle \phi^r \rangle =0$.
  (We have divided the gauge index $a = (i, r)$ as $i$ and $r$ label the Cartan and non-Cartan parts respectively.)
  Extremizing the scalar potential, we obtain the following conditions 
    \bea
    \langle
    \CF_{ade} g^{bd} g^{ec} 
    (e \delta^0_{b} + m \bar{\CF}_{0b})(\bar{e} \delta^0_{c} + m \bar{\CF}_{0c})
    \rangle
     =    0.
    \eea
  We have denoted the derivative of $\CF$ with respect to $\phi_a, \phi_b, \ldots$ as $\CF_{a b \ldots}$. 
  
  In order to analyze the above conditions, let us change the basis for Cartan part 
  such that the Cartan generators in the new basis are 
  $(t_{\ui})_j^{~k} = \delta_{\ui}^{~k} \delta_j^{~\ui}$ ($\ui = 1, \ldots, N$) \cite{FIS2}.
  In this basis, the conditions of the vacua are simply written as
    \bea
    \langle
    \CF_{\ui \ui \ui} (g^{\ui \ui})^2 (2 e + m \bar{\CF}_{\ui \ui})(2 \bar{e} + m \bar{\CF}_{\ui \ui})
    \rangle
     =      0,
            ~~~~
            \ui~{\rm not ~summed},
            \label{cond}
    \eea
  for each $\ui$. 
  Since $\langle \CF_{\ui \ui \ui} (g^{\ui \ui})^2 \rangle = 0$ corresponds to unstable vacua, 
  the above condition reduces to 
  $\langle (2 e + m \bar{\CF}_{\ui \ui})(2 \bar{e} + m \bar{\CF}_{\ui \ui}) \rangle = 0$ for each $\ui$.
  The parameter $e$ is complex, thus we have to choose 
  $\langle \CF_{\ui \ui} \rangle = - 2e/m$ or $- 2 \bar{e}/m$ for each $\ui$.
  In terms of the K\"ahler metric, these become the conditions
    \bea
    \langle g_{\ui \ui} \rangle
     =   - \frac{2 \Im e}{m}
           ~{\rm or}~
           \frac{2 \Im e}{m},
    \eea
  for each $\ui$.
  It should be noted that 
  these $\langle g_{\ui \ui} \rangle$ correspond to the squared gauge coupling constant $(1/g^2)_{\ui}$.
  Therefore, we should be careful when $\langle g_{\ui \ui} \rangle$ are negative for some $\ui$.
  
  Taking this into consideration, we now classify the vacua. 
  Let $N_+$ denote the number of $\ui$ such that $\langle g_{\ui \ui} \rangle = - 2 \Im e/m$.
  In the case with $\Im e/m < 0$, we can divide the vacua into the following three groups:
    \begin{enumerate}
      \item $N_+ = N$. 
            In this case, we have $\langle g_{\ui \ui} \rangle = - 2 \Im e/m$ for all $\ui$.
            These vacua correspond to what are obtained from the F-term condition, 
            that is, these vacua preserve the $\CN=1$ supersymmetry 
            which is manifest in the Lagrangian (\ref{lagrangian}).
            Also, the vev of K\"ahler metric determines the vev of the scalar field.
            According to this, the gauge symmetry is generally broken as
              \bea
              U(N) 
               \rightarrow
                     \prod_{i=1}^n U(N_i).
                     \nonumber
              \eea 
      \item $N_+ = 0$.
            In this case, $\langle g_{\ui \ui} \rangle = 2 \Im e/m < 0$ for all $\ui$.
            In contrast to the vacua 1, these vacua preserve another $\CN=1$ supersymmetry, 
            as we will see in section \ref{subsec:partial}.
            Since $\langle g_{\ui \ui} \rangle$ or $(1/g^2)_{\ui}$ are all negative in these vacua,
            it seems that these vacua have no meaning.
            However, by applying the analysis of the last section, 
            we can propose a good description of this.
            We will see this in the subsequent sections.
      \item $N_+ \neq N$ and $N_+ \neq 0$.
            In these cases, we can set $\langle g_{\ui \ui} \rangle = - 2 \Im e/m > 0$ for $\ui = 1, \ldots, N_+$
            and $\langle g_{\ui \ui} \rangle = 2 \Im e/m < 0$ for $\ui = N_+ + 1, \ldots, N$
            by appropriate Weyl reflection.
            These vacua break both $\CN=1$ supersymmetries.
            We will check this below.
            So, we will not consider these vacua any more in this paper.
    \end{enumerate}
  The most striking point of this model is that the vacua 2 preserve the second $\CN=1$ supersymmetry, 
  though the squared gauge coupling constants $(1/g^2)_{\ui}$ are negative in these vacua.
  This is because this model has partially broken $\CN=2$ supersymmetry.
  Note that, in the generic model, we can also analyze the scalar potential as well.
  Then, we can obtain some vacua 
  in addition to the supersymmetric vacua obtained from the F-term equation.
  However, this does not lead to interesting result
  because these vacua are merely non-supersymmetric vacua.
  In contrast to this, the model here has the supersymmetric vacua 2 in addition to the vacua 1.
  
  In the case with $\Im e/m > 0$, the situation is opposite to the above.
  We also divide the vacua into the similar three types:
    \begin{enumerate} \setcounter{enumi}{3}
      \item $N_+ = N$. 
            In this case, we have $\langle g_{\ui \ui} \rangle = - 2 \Im e/m < 0$ for all $\ui$.
            Although these vacua preserve the manifest $\CN=1$ supersymmetry as well as the vacua 1, 
            the squared gauge coupling constants are negative.
      \item $N_+ = 0$. 
            $\langle g_{\ui \ui} \rangle = 2 \Im e/m > 0$ for all $\ui$.
            As the vacua 2, these vacua preserve the second $\CN=1$ supersymmetry.
      \item $N_+ \neq N$ and $N_+ \neq 0$.
            These vacua break both $\CN=1$ supersymmetries.
            We will not consider these here.
    \end{enumerate}
  
  Finally, let us make a comment on the vacuum energy.
  It is straightforward to compute this from the expression of the scalar potential (\ref{potential}):
  it is zero (or $8N m \Im e$) in the vacua 1 and 4 (or in the vacua 2 and 5).
  Note that non-zero vacuum energy does not mean that the supersymmetry is broken in the vacuum.
  In fact, we have a freedom to shift vacuum energy by a constant.
  From the $\CN=2$ perspective, this corresponds to the different fixing of the electric FI parameters
  when we fix them \cite{APT, Fujiwara, IMS, OOP}.
  Here we have fixed them such that the vacuum energy of the vacua which preserve the manifest $\CN=1$ supersymmetry 
  becomes zero.

\subsection{Partial supersymmetry breaking}
\label{subsec:partial}
  We now examine supersymmetry preserved in the above vacua.
  As noted above, we can easily see that the vacua 1 and 4 are supersymmetric 
  because $\langle g_{\ui \ui} \rangle = - 2 \Im e/m$ are derived from the F-term condition $\partial_\Phi W =0$.
  We can also check this and, furthermore, see the unbroken supersymmetry in the vacua 2 and 5
  by observing the appearance of a Nambu-Goldstone fermion.
  
  For this purpose, let us consider the supersymmetry transformation law of the fermions, 
  which is written, in the eigenvalue basis, as
    \bea
    \delta \left(
    \begin{array}{c}
    \lambda^{\ui} \\
    \psi^{\ui}
    \end{array}
    \right) 
     =   - \sqrt{2} g^{\ui \ui}
           \left(
           \begin{array}{cc}
           0 & -( 2 e + m \bar{\CF}_{\ui \ui}) \\
           2 \bar{e} + m \bar{\CF}_{\ui \ui} & 0
           \end{array}
           \right)   
           \left(
           \begin{array}{c}
           \eta_1 \\
           \eta_2
           \end{array}
           \right)     
         + \ldots,
           \eea
  where we have omitted the terms which are trivially zero in the vacuum.
  $\eta_1$ and $\eta_2$ are the supersymmetry transformation parameters 
  of the $\CN=1$ supersymmetry manifest in the Lagrangian (\ref{lagrangian})
  and the second $\CN=1$ supersymmetry respectively.
  
  In the case with $N_+ = N$, that is, $\langle \CF_{\ui \ui} \rangle = - 2 e/m$ for all $\ui$, 
  the transformation law at these vacua becomes
    \bea
    \langle \delta \lambda^{\ui} \rangle
     =   - 2 \sqrt{2} m i \eta_2,
           ~~~
    \langle \delta \psi^{\ui} \rangle
     =     0,
           ~~~
           {\rm for~all~}\ui.
    \eea
  If we change the basis into the original Cartan basis, we obtain
    \bea
    \langle \delta \lambda^a \rangle
     =   - 4 \sqrt{N} m i \eta_2 \delta_0^a,
           ~~~
    \langle \delta \psi^a \rangle
     =     0,
           \label{2break}
    \eea
  where a factor $\sqrt{N}$ comes from the convention $t_0 = 1_{N \times N}/\sqrt{2N}$.
  Therefore, it follows that the second $\CN=1$ supersymmetry is broken 
  and manifest $\CN=1$ supersymmetry is preserved in these vacua.
  Also, $\lambda^0$ is a Nambu-Goldstone fermion.
  On the other hand, when $N_+ = 0$, that is, $\langle \CF_{\ui \ui} \rangle = - 2 e/m$ for all $\ui$,
  the transformation law becomes in the original basis as follows:
    \bea
    \langle \delta \lambda^a \rangle
     =     0,
           ~~~
    \langle \delta \psi^a \rangle
     =     4 \sqrt{N} m i \eta_1 \delta^a_0.
           \label{1break}
    \eea
  We can see that the manifest $\CN=1$ supersymmetry is broken, but the second $\CN=1$ supersymmetry is preserved
  and $\psi^0$ is a Nambu-Goldstone fermion.
  The fact that the above vacua preserve $\CN=1$ supersymmetry can also be seen
  by examining the mass spectrum and observing the massless Nambu-Goldstone fermion as in \cite{FIS2}.
  In fact, we can see that the fermion $\lambda$ (or $\psi$) which includes the Nambu-Goldstone fermion
  makes $\CN=1$ vector multiplet with the gauge field 
  and the fermion $\psi$ (or $\lambda$) is combined with the scalar field to be $\CN=1$ massive chiral multiplet.
  
  In the other cases with $N_+ \neq N$ and $N_+ \neq 0$, we have
    \bea
    \langle \delta \lambda^{\ui} \rangle
    &=&  - 2 \sqrt{2} m i \eta_2,
           ~~~
    \langle \delta \psi^{\ui} \rangle
     =     0,
           ~~~
           {\rm for}~\ui = 1, \ldots, N_+,
           \nonumber \\
    \langle \delta \lambda^{\ui} \rangle
    &=&    0,
           ~~~
    \langle \delta \psi^{\ui} \rangle
     =     2 \sqrt{2} m i \eta_1,
           ~~~
           {\rm for}~\ui = N_+ + 1, \ldots, N.
    \eea
  Thus, there are two Nambu-Goldstone fermions in the model and 
  $\CN=2$ supersymmetry is broken completely.
  We have, therefore, checked the unbroken supersymmetry preserved in various vacua.
  
\subsection{Large FI parameters limit}
  The models considered in this paper are the extensions 
  of $\CN=1$, $U(N)$ super Yang-Mills theory with the tree level superpotential.
  As we can recover this from the generic model considered in section \ref{sec:duality} 
  by taking the parameters $t_k$ ($k>0$) to zero, 
  the model considered in this section also reduces to $\CN=1$, $U(N)$ super Yang-Mills theory with the superpotential
  by the limit where $e, m \rightarrow \infty$ with $a_k = m g_k$ ($k \geq 2$) and $e/m$
  fixed \cite{FIS, Fujiwara}.
  We refer to this limit as large FI parameters limit.
  
  In the large FI parameters limit where $m \rightarrow \infty$, 
  the dominant part in the supersymmetry transformation law of the fermion is indeed the term in
  (\ref{2break}) (or (\ref{1break})). 
  Therefore the broken supersymmetry leads to the fermionic symmetry 
  which acts the field strength superfields as $\CW^{i \alpha} \rightarrow \CW^{i \alpha} + \eta 1_{N_i \times N_i}$.
  Note that parameter $\eta$ is related to the broken supersymmetry parameter $\eta_2$ (or $\eta_1$) 
  appeared in (\ref{2break}) (or (\ref{1break})).
  This is the fermionic shift symmetry which has been argued in \cite{CDSW}.
  
  One may consider that the vacua 2 and 5 which preserve the second $\CN=1$ supersymmetry are inconsistent with 
  the vacuum structure of $\CN=1$, $U(N)$ super Yang-Mills theory with the superpotential.
  However, the vacua 2 (or 5) are decoupled from the vacua 1 (or 4) in this limit.
  This can be seen from the vacuum condition (\ref{cond}).
  For example, we first manage to keep the vev of $\phi$ in the vacua 1 (or 4) at finite value, 
  by setting $2 e + m g_1 = 0$.
  In this case of the parameters, the vev of $\phi$ in the vacua 2 (or 5) becomes infinite in the limit 
  and these vacua are decoupled from each other.
  We can also observe this from the simple analysis of the vacuum energy of each type of the vacua.
  As we saw above, the vacuum energy of the vacua 1 (or 4) is zero.
  On the other hand, that of the vacua 2 (or 5) is proportional to $m \Im e$ and this becomes infinite in the limit.
  Therefore, in this limit, the vacua 1 (or 4) are far away from the vacua 2 (or 5) 
  and we can only see either of them.
  This matches with the vacuum structure of $\CN=1$, $U(N)$ super Yang-Mills theory with the superpotential.
     
\section{Non-supersymmetric vacua}
\label{sec:non-susy}
  So far, we have analyzed the classical vacua.
  In this section, we will consider the meaning of the vacua 2 and 4
  and we will see that these lead to non-supersymmetric vacua by using SUSY/non-SUSY duality.
  Therefore, we can see that there are supersymmetry breaking vacua 
  in addition to the supersymmetric vacua of the effective theory. 
  We will analyze these by using the different gluino condensate effective superpotential 
  written in terms of $\theta_1$ and $\theta_2$.
  (We will use $\theta_1$ and $\theta_2$ to denote the superspace coordinate of the manifest $\CN=1$ supersymmetry
  and the second $\CN=1$ supersymmetry respectively.)
  
  First of all, let us consider the case with $\Im e/m <0$ and concentrate on the vacua 1 in the classical theory.
  These vacua preserve the manifest $\CN=1$ supersymmetry and the gauge symmetry is broken to $\prod_i U(N_i)$.
  Supposing that each $SU(N_i)$ factor confines at the low energy, we can evaluate the effective superpotential
  in terms of the gluino condensate superfields
    \footnote{We write down the effective superpotential in terms of $S_i$, 
              where these are defined as the $U(N_i)$ traces of the squared field strength superfields.
              Thus, $S_i$ here are slightly different from the gluino condensate fields 
              which are the $SU(N_i)$ trace of them.}.
  This can be obtained by the supergraph technique and the diagrammatical computation \cite{IM} 
  or by using the generalized Konishi anomaly equations and the chiral ring property \cite{IM, IM2} as follows:
    \bea
    \int d^2 \theta_1 W_{{\rm eff}},
           ~~~~
    W_{{\rm eff}}
     =     \sum_i N_i \frac{\partial F_{{\rm free}}}{\partial S_i}~
        +  \frac{16 \pi^2 i }{m}
           \left(
        -  2 e \sum_i S_i
        +  \sum_{k = 2}^{n+1} g_k~
           \frac{\partial F_{{\rm free}}}{\partial g_{k -1}}
           \right),
           \label{weff1}
    \eea
  where $F_{{\rm free}}$ is the free energy of the bosonic one matrix model
  whose action is the same form as the superpotential in (\ref{prepot}).
  The supersymmetric vacua can be derived 
  from the F-term condition $\partial_{S_i} W_{{\rm eff}} = 0$.
  
  Note that the last terms disappear in large FI parameters limit 
  where $m g_k = a_k$ (for $k \geq 2$) are fixed, 
  because $F_{{\rm free}}$ depends only on $a_k$ and not on $m$ \cite{IM2}.
  Therefore, we obtain the effective superpotential considered by Dijkgraaf and Vafa \cite{DV}
  as a particular limit of (\ref{weff1}).
  
  Now, let us flip the sign of the FI parameter $\Im e$ and go to the region 
  where the squared gauge coupling constants are negative in the classical theory.
  In this case, the structure of the classical vacua of the model becomes the latter pattern
  in section \ref{subsec:vacua}, that is, the vacua 4, 5 and 6.
  The effective superpotential with this sign of $\Im e$, 
  which is written in terms of the superspace $\theta_1$, may denote that of the vacua 4 in the classical theory
  where the squared gauge coupling constants are negative.
  According to the suggestion in section \ref{sec:duality}, there is, however, a better description in UV of these,
  which is $U(N)$ gauge theory with the following gauge kinetic term and a tree superpotential
    \bea
    \int d^2 \tilde{\theta}_1 
    \Tr \left[
    - \frac{i}{4} \left(- \frac{2 \bar{e}}{m} \right) \widetilde{\CW}^\alpha \widetilde{\CW}_\alpha
    + W (\widetilde{\Phi}) \Big|_{g_k \rightarrow g_k + 16 \pi^2 i g_{k+1} \tilde{\theta}_1^2/m}
    \right],
    \label{dual1}
    \eea
  where $\tilde{\theta}_1$ is the different from the manifest $\CN=1$ superspace coordinate $\theta_1$ 
  and the tilded superfields are defined on $\tilde{\theta}_1$ superspace.
  Also, the form of the superpotential in (\ref{dual1}) is the same as that of the original one in (\ref{prepot}).
  Note that $e$ has been changed to $\bar{e}$ when we have gone to the dual description, 
  in order for the squared gauge coupling constants to be positive in the dual theory.
  As we saw above, this is supersymmetry breaking model 
  by the spurion fields which get the vevs in the auxiliary components.
  
  So far, we have only considered the effective theory in terms of $\theta_1$ coordinate
  and the analysis was the similar as that in section \ref{sec:duality}.
  However, the most crucial point in this model is the existence of the second $\CN=1$ supersymmetry
  in the classical theory, in other words, the existence of the vacua 2 or 5 as we saw in the last section.
  We cannot analyze these vacua from the above gluino condensate effective superpotential 
  which is written in terms of $\theta_1$ coordinate, 
  because $\theta_1$ is the different coordinate from that of the $\CN=1$ supersymmetry 
  preserved in the vacua 2 and 5.
  Instead, we will analyze the gluino condensate effective superpotential written in terms of $\theta_2$.
  
  In order to analyze the vacua 2, we start from the vacua 5 by setting $\Im e/m$ to be positive.
  Since these vacua have non-zero vacuum energy as we saw in section \ref{subsec:vacua}, we rewrite the F-term part 
  of the classical scalar potential (\ref{potential}) as
    \bea
    g^{ab} \partial_a W_2 (\Phi) \overline{\partial_b W_2 (\Phi)} + 8N m \Im e, 
    \label{newpot}
    \eea
  where the \textit{new} superpotential $W_2(\Phi)$ is defined by
    \bea
    W_2 (\Phi)
     =     2 \bar{e} \Phi + m \frac{\partial \CF}{\partial \Phi}.
           \label{w2}
    \eea
  In this convention, we can manifestly see the second $\CN=1$ supersymmetry, 
  because the F-term condition $\partial_\Phi W_2 (\Phi) =0$ leads to $\langle g_{\ui \ui} \rangle = 2 \Im e/m$,
  the condition of the vacua which preserve the second $\CN=1$ supersymmetry.
  By using these, we can evaluate the effective superpotential in a similar way 
  except that the gluino condensate field should be different from $S_i$ in the above case. 
  If we denote these as $T_{i}$, the effective superpotential is
    \bea
    \int d^2 \theta_2 W_{{\rm eff}} (T),
           ~~~~
    W_{{\rm eff}}
     =     \sum_i N_i \frac{\partial F_{{\rm free}}}{\partial T_{i}}~
        +  \frac{16 \pi^2 i }{m}
           \left(
        -  2 \bar{e} \sum_i T_i
        +  \sum_{k = 2}^{n+1} g_k~
           \frac{\partial F_{{\rm free}}}{\partial g_{k -1}}
           \right),
           \label{weff2}
    \eea
  where $F_{{\rm free}}$ is the same as the one in (\ref{weff1}) but it is the function of $T_i$.
  The supersymmetric vacua are derived from F-term condition as above.
  Also, it should be noted that the vacuum energy is not zero, but $8N m \Im e$ 
  as appeared in the classical potential (\ref{newpot}).
  
  We want to understand the vacua 2 where the squared gauge coupling constants are negative.
  In order to do, we propose a better UV description of these vacua,
  which is 
    \bea
    \int d^2 \tilde{\theta}_2 
    \Tr \left[
    - \frac{i}{4} \left(- \frac{2 e}{m} \right) \widetilde{\CW}^\alpha \widetilde{\CW}_\alpha
    + W_2 (\widetilde{\Phi}) \Big|_{g_k \rightarrow g_k + 16 \pi^2 i g_{k+1} \tilde{\theta}_2^2/m}
    \right].
    \label{dual2}
    \eea
  Note that the superpotential of the dual description is not same as the original form (\ref{prepot}),
  but has the same form as (\ref{w2}).
  Also, we can estimate the vacuum energy of these vacua.
  The above shifts in the superpotential contribute to the potential as $V \sim - \Im e /m$.
  In addition, we should add $8N m \Im e$ which is the vacuum energy of the original theory (\ref{newpot}) to this.
  
  In \cite{GMT}, the effective superpotential of the model where the supersymmetry is broken 
  by the vev of the auxiliary fields of the spurion superfields has been derived.
  If we apply this to our case, 
  the effective superpotential of the dual description (\ref{dual2}) can be written by the following simple formula:
    \bea
    \int d^2 \tilde{\theta}_2 \widetilde{W}_{{\rm eff}},
    ~~~
    \widetilde{W}_{{\rm eff}}
     =   - \frac{32 \pi^2 i e}{m} \sum_i \widetilde{T}_i
         + N_i \frac{\partial F_{{\rm free}}}{\partial \widetilde{T}_i}
           \Big|_{g_k \rightarrow g_k + 16 \pi^2 i g_{k+1} \tilde{\theta}_2^2/m}.
           \label{dualweff2}
    \eea
  As explained in \cite{ABSV}, an evidence of the duality can be seen from these effective superpotential
  (\ref{weff2}) and (\ref{dualweff2}), by using the $\CN=2$ structure of the low energy effective theory:
  (\ref{weff2}) can be obtained from $\int d^2 \theta_2 d^2 \tilde{\theta}_2 F_{{\rm free}}$
  by shifting $T_i \rightarrow T_i +  N_i \tilde{\theta}^2_2$ and 
  $g_k \rightarrow g_k + 16 \pi^2 i g_{k+1} \tilde{\theta}^2_2/m $
  and integrating out $\tilde{\theta}_2$.
  Also, (\ref{dualweff2}) is obtained from the same one by $T_i \rightarrow T_i + N_i \theta_2^2$ and 
  $g_k \rightarrow g_k + 16 \pi^2 i g_{k+1} \tilde{\theta}^2_2/m $
  and integrating out $\theta_2$.
  The only difference is the shift of $T_i$ and is an auxiliary field redefinition.
  Therefore, these describe the same physics.
  
  In principle, we can obtain the value of $\langle S_i \rangle$ by extremizing the scalar potential 
  which computed from (\ref{dualweff2}) or (\ref{weff2}).
  In such vacua, the supersymmetry is broken as trivially seen from (\ref{dualweff2}).
  Note that this is the case with $\Im e /m <0$ and there are the supersymmetric vacua which derived from (\ref{weff1})
  in addition to these supersymmetry breaking vacua.
  
  It should be noted that we have used various superspace coordinates: 
  $\theta_2$ (the superspace coordinate of the second supersymmetry), 
  $\tilde{\theta}_1$ (the coordinate where the Lagrangian (\ref{dual1}) of the dual description is defined)
  and $\tilde{\theta}_2$ (the coordinate where (\ref{dual2}) is defined).
  Since the origins of the dual coordinates $\tilde{\theta}_1$ and $\tilde{\theta}_2$ are in 
  $\CN=2$ structures of the effective theories, so it seems that some of them are related. 
  For example, one may consider that $\tilde{\theta}_2$ and $\theta_1$ are the same coordinates.
  However, it is difficult to verify this, 
  because the effective theories are written in terms of the different gluino condensate superfields.
  Instead, we can see that at least, in the large FI parameters limit, they are quite different coordinates.
  Let us see this here, concentrating on $\tilde{\theta}_2$ and $\theta_1$.
  Although it seems that the holomorphic part of the original Lagrangian written in terms of $\theta_1$ and 
  dual description (\ref{dual2}) written in terms of $\tilde{\theta}_2$ become the similar form in this limit, 
  the signs of $\Im e$ in the superpotential are different.
  Thus, they describe the different theories.
  This can be also seen from the vacuum energy: the vacuum energy of the vacua 1 is exactly zero. 
  On the other hand, as we have seen above, that of the theory (\ref{dual2}) is not zero even in the limit.
  Actually, it is natural to consider that these vacua are decoupled each other
  and in $\CN=1$, $U(N)$ super Yang-Mills theory with superpotential, 
  we can observe only either of them, as we have seen in section \ref{subsec:vacua}.
  Also, we can see that $\theta_2$ and $\tilde{\theta}_1$ are quite different in the limit as well.
  
  We can say that the effective theory has the supersymmetric vacua and also the supersymmetry breaking vacua.
  Note that this does not mean 
  that, in the generic model, there are either of the supersymmetric vacua and the supersymmetry breaking vacua.
  The generic model may have the similar vacua as above.
  However, we do not have the way to analyze both in the generic model 
  because it has only $\CN=1$ supersymmetry at the classical level.
  
\section{Summary and discussion}
\label{sec:summary}
  We have analyzed the vacua of the $U(N)$ gauge model with partially broken $\CN=2$ supersymmetry.
  We here summarize the results, according to the classification in section \ref{subsec:vacua}.
  In the case with $\Im e /m < 0$, we have seen that the classical vacua are divided into the three types
  (we have not considered the third type vacua):
    \begin{enumerate}
      \item In these vacua, the manifest $\CN=1$ supersymmetry is preserved classically.
            (We have denoted the superspace coordinate of this as $\theta_1$.)
            We have analyzed the effective theory from the effective superpotential 
            and have seen the (manifest) $\CN=1$ supersymmetric vacua at low energy.
      \item In this case, the classical vacua preserve the second $\CN=1$ supersymmetry, 
            but all the squared gauge coupling constants are negative.
            Based on the proposal in \cite{ABSV}, we have proposed a dual description of these in UV,
            which is the different $U(N)$ gauge model with (\ref{dual2}), 
            written in terms of superspace $\tilde{\theta}_2$.
            We have seen the effective superpotential of such a theory
            and the existence of supersymmetry breaking vacua.
            The existence of the classical vacua 2 which preserve the second $\CN=1$ supersymmetry
            is the crucial point of this model and this has made the above analysis possible.
    \end{enumerate}
  We have also seen that these superspace coordinates $\theta_1$ and $\tilde{\theta}_2$ are different each other.
  In the case with $\Im e/m > 0$, we can do the same analysis:
    \begin{enumerate} \setcounter{enumi}{3}
      \item In this case, the squared gauge coupling constants are negative classically.
            We can analyze this case by SUSY/non-SUSY duality as in \cite{ABSV}
            and the dual description in UV is the explicit supersymmetry breaking model as above.
      \item As the vacua 2, these vacua preserve the second $\CN=1$ supersymmetry classically.
            We have analyzed the effective superpotential 
            and we have seen the second $\CN=1$ supersymmetric vacua at low energy.
            These \textit{supersymmetric} vacua do not exist in the generic model and 
            the existence of these is also the remarkable property of this model.
    \end{enumerate}
  
  It is interesting to consider this model from the string theory perspective.
  The generic model has already been studied in type IIB superstring theory \cite{ABSV} 
  and in type IIA and M-theory \cite{OT}.
  However, it is not so clear why the symmetry becomes higher 
  in this specific choice of the parameters from the IIB superstring theory point of view.

\section*{Acknowledgements}
  The author thanks Kazuo Hosomichi, Hironobu Kihara, Sungjay Lee, 
  Alberto Mariotti, Kazutoshi Ohta, Yutaka Ookouchi, Takao Suyama, Yuji Tachikawa, Masato Taki, Futoshi Yagi 
  and especially Hiroshi Itoyama
  for useful discussions and comments.
  The author also thanks Korea Institute for Advanced Study for the hospitality during part of this project.
  The research of the author is supported in part by JSPS Research Fellowships for Young Scientists.




\end{document}